\documentstyle[aps,prl]{revtex}
\begin{document}
\draft
\twocolumn[
\title{Electronic states on a twin boundary of a $d$-wave superconductor}
\author{M. E. Zhitomirsky\cite{Landau} and M. B. Walker}
\address{Department of Physics, University of Toronto, Toronto, 
Canada M5S 1A7}
\date{May 9, 1997}
\maketitle

\widetext \advance\leftskip by 57pt \advance\rightskip by 57pt
\begin{abstract}
We show that an induced $s$-wave harmonic in the superconducting gap 
of an orthorhombic $d_{x^2-y^2}$ superconductor strongly affects the 
excitation spectrum near a twinning plane. In particular, it yields 
bound states of zero energy with areal density proportional to the 
relative weight of the $s$-wave component. An unusual scattering 
process responsible for the thermal conductivity across the twin 
boundary at low temperatures is also identified. 
\end{abstract}
\pacs{PACS numbers:
    74.25.Jb,  
    74.25.Fy,  
    74.72.Bk   
}
]\narrowtext

The controversy concerning $s$-wave versus $d$-wave pairing in 
high-$T_c$ superconductors is being gradually resolved in favor of the 
$d_{x^2-y^2}$ symmetry. This is possible because of precise 
measurements of spontaneously generated half-integral flux quanta on 
bicrystal and tricrystal films \cite{halfquant} together with the 
corner SQUID experiments \cite{corner}, which all suggest that the 
superconducting gap changes sign on the Fermi surface between the $a$- 
and $b$-directions with zeros or nodes in between. The existence of 
corresponding low-energy excitations in the superconducting state of 
cuprates have also been demonstrated by measurements of the magnetic 
penetration depth \cite{penetdepth}, the nuclear spin relaxation rate 
\cite{NMR}, and by angular resolved photoemission data \cite{ARPES}. 

However, not all experimental data fit into a simple picture of a pure 
$d_{x^2-y^2}$ order parameter. Sun {\it et al\/}.\ \cite{caxis} have 
found a non-vanishing tunneling current along the $c$-axis between 
YBa$_2$Ca$_3$O$_{6+x}$ (YBCO) and the conventional superconductor Pb, 
which cannot exist for a tetragonal $d$-wave superconductor. It was 
argued \cite{Sigrist96,Walker96} that such an experiment can be 
understood if an admixture of an $s$-wave component due to the 
orthorhombicity of YBCO crystals \cite{ortho} is taken into account. 
The mixed wave $d+s$ gap then has nodes on the Fermi surface shifted 
from the principal diagonal directions. Recent high resolution 
measurements of the $ab$-plane thermal conductivity have failed to see 
this effect imposing an upper limit of 10\% on the weight of the 
$s$-wave harmonic \cite{thermcond}. 

An important challenge for current research is to reconcile various 
conflicting results on symmetry of the superconducting order parameter 
in YBCO by direct determination of the $s$-wave component in the gap. 
In this work we study the structure of the quasiparticle spectrum 
close to a twinning plane in a predominantly $d$-wave orthorhombic 
superconductor and show that it changes significantly in the presence 
of an $s$-wave component of the gap. In particular, there appear bound 
states of zero energy with areal density proportional to the magnitude 
of the $s$-wave component. Bound states cause a modification of the 
local density of states which can be observed by scanning tunneling 
microscopy. Other new effects studied below include an unusual low 
temperature mechanism of heat transport across the twin boundaries. 
This is a kind of Andreev transmission process because it involves 
particle-hole conversion. 

The tetragonal symmetry of the CuO$_2$ planes in YBCO is spoiled by 
the presence of CuO chains. The two possible orientations of the 
chains lead to the formation of so-called twin domains separated by 
twin boundaries (TBs), which are (110) or $(1\bar{1}0)$ planes of the 
original tetragonal lattice (Fig~\ref{TB}). The structural anisotropy 
gives rise to an anisotropy in the pairing interaction and, hence, to 
an admixture (real combination) of $s$- and $d$-waves in the 
superconducting gap with the relative phase $0$ or $\pi$ in different 
twins \cite{ortho}. Since we assume the dominant component of the gap 
to be of the $d$-wave symmetry, a natural energy requirement is that 
$\psi_d$ stays almost constant across the boundary, whereas $\psi_s$ 
has different signs on both sides, thus exhibiting a soliton-like 
structure. The symmetry of the combined order parameter ($d\pm s$) is, 
in this case, odd with respect to the reflection in the twinning 
plane. Note, that the opposite assumption of an even symmetry state 
($s\pm d$) is in  conflict with the $ab$-tunneling experiments on 
heavily twinned YBCO samples \cite{Walker96}. Sigrist and co-workers 
\cite{Sigrist96} suggested that the $s$-wave component could change 
sign between two twin domains by avoiding zero at the twin boundary 
and forming a time-reversal breaking state $d+se^{i\chi}$. Existence 
of the time-reversal breaking states on twin boundaries in YBCO is 
still an open question, we therefore leave discussion of the 
electronic spectrum for such a twin boundary to the end. 

The soliton-like behavior of the induced $s$-wave component plays a 
key role in the formation of bound states on twin boundaries. Examples 
of fermions trapped on one-dimensional inhomogeneities of the 
off-diagonal potential have been known for a long time 
\cite{polyacyt}. To get some insight into this we first consider a 
very simplified model of the boundary between two twins. Namely, we 
neglect all orthorhombic effects except the induced inhomogeneous 
$s$-wave component and assume cylindrical Fermi surface. The 
Bogoliubov-de Gennes (BdG) equations in the quasiclassical 
approximation \cite{Bruder90}, i.e., retaining terms of the lowest 
order in $(k_F\xi_0)^{-1}$, are written for a slowly varying function 
of space coordinates $(u,v)=e^{-i{\bf k}_F\cdot{\bf r}}\psi$, $\psi$ 
being the usual BdG wave function, as 
\begin{equation}
E \left(\begin{array}{c}u\\v\end{array}\right) =
\left( \begin{array}{cc}          
-i{\bf v}_F\!\cdot\!\mbox{\boldmath$\nabla$}&\Delta({\bf k}_F,{\bf r})\\
\Delta^*({\bf k}_F,{\bf r})&i{\bf v}_F\!\cdot\!\mbox{\boldmath$\nabla$}
\end{array}\right) 
\left(\begin{array}{c} u \\ v \end{array}\right) , 
\label{Andreev}
\end{equation} 
where ${\bf v}_F={\bf k}_F/m$ ($\hbar=1$). For the pair potential 
close to the twin boundary ($x=0$ plane) we choose 
\begin{equation}
\Delta({\bf k},{\bf r}) = \Delta_d({\bf k})+\Delta_s\tanh(x/\xi_0) \ ,
\label{ansatz}
\end{equation}
which gives correct orthorhombic gaps $\Delta_d\pm\Delta_s$ at 
$x\rightarrow\pm\infty$. At $E=0$, system (\ref{Andreev}) can be 
transformed to a pair of independent first-order differential 
equations, which are easily solved giving 
\begin{eqnarray}
\psi_+ & = & C e^{i{\bf k}_F\cdot{\bf r}} e^{-x/\xi_d} 
  \frac{1}{[\cosh(x/\xi_0)]^{\xi_0/\xi_s}} 
\left(\begin{array}{c}1\\-i\end{array}\right) \nonumber \\
\psi_- & = & C e^{i{\bf k}'_F\cdot{\bf r}} e^{x/\xi_d}
  \frac{1}{[\cosh(x/\xi_0)]^{\xi_0/\xi_s}} 
\left(\begin{array}{c}1\\i\end{array}\right) ,
\label{states}
\end{eqnarray}
where ${\bf k}'_F=(-k_{Fx},k_{Fy})$, $\Delta_d({\bf k}'_F) = 
-\Delta_d({\bf k}_F)$, $\xi_d=v_{Fx}/|\Delta_d({\bf k}_F)|$, 
$\xi_s=v_{Fx}/\Delta_s$. These solutions correspond to physical states 
whenever normalization condition can be satisfied. Since the 
asymptotic behavior of $\psi$ does not depend on $\xi_0$, we 
immediately conclude that the bound state for a given ${\bf k}_F$ 
exists if $\Delta({\bf k}_F,\infty)$ and $\Delta({\bf k}_F,-\infty)$ 
have {\it different} signs, e.g., for $k_{y\text{C}'} < k_{Fy} < 
k_{y\text{A}}$ in Fig.~\ref{FS}. This requirement can also be written 
as $|\Delta_d({\bf k}_F)|<\Delta_s$. For a model $d$-wave gap 
$\Delta_d({\bf k}) = \Delta_0(k_a^2-k_b^2)/k_F^2$ or, in our 
coordinate system, $\Delta_d({\bf k})=2\Delta_0k_xk_y/k_F^2$, the 
midgap excitations exist in the vicinity of $(\pm k_F,0)$ for 
$|k_y|<(\Delta_s/2\Delta_0)k_F$ and close to $(0,\pm k_F)$ with the 
same condition on $k_x$, though in the latter case, as we shall see 
below, these states are easily destroyed by various perturbations. 

The local density of states near the twin boundary has a $\delta$-like 
peak at $E=0$ corresponding to the bound states. The associated areal  
density 
\begin{equation} 
N_B(E) = \delta(E)\sum_n \int |v^b_n(x)|^2 dx = 
\frac{2k_F}{\pi}\frac{\Delta_s}{\Delta_0}\delta(E)
\label{locden}
\end{equation} 
is proportional to the relative weight of the $s$-wave component. 
Furthermore, it is possible to show with the help of the Atiyah-Singer 
theorem that the zero-energy branch found for the gap function 
(\ref{ansatz}) exists for a wide class of gaps between two twins 
\cite{AS}. 

To describe more realistically anisotropy of electron properties in 
YBCO we now consider a second model, which is a tight-binding model of 
a single CuO$_2$ plane. The model includes the orthorhombicity through 
unequal effective hopping matrix elements $t_1$ and $t_2$ along the 
$a$ and $b$ axes. The two different hopping amplitudes allow us to 
construct a twinning plane on a square lattice as shown in 
Fig.~\ref{TB}. An extra local potential (which we denote by $U_0$ and 
assume positive) associated with atoms on the twin boundary is allowed 
by the symmetry and will be shown to have important consequences for 
the twin boundary properties. Thus, the one-electron Hamiltonian of 
the normal phase is 
\begin{equation}
\hat{H}_0 = \sum_{ij\sigma}(t_{ij}-\mu\delta_{ij})
c_{i\sigma}^\dagger c_{j\sigma}^{_{}}+U_0\sum_{i\in\text{TB}\sigma}
c^\dagger_{i\sigma}c_{i\sigma}^{_{}} \ .
\label{H}
\end{equation}
The Schr\"odinger equation for the discrete function 
$\psi_{i\sigma}$ is 
\begin{equation}
(E+\mu)\psi_{i\sigma} = \sum_jt_{ij}\psi_{j\sigma} + 
 \left. U_0 \psi_{i\sigma}\right|_{i\in\text{TB}} \ .
\label{Sch}
\end{equation}
The dispersions of plane waves in the two twins are 
\begin{eqnarray}
\varepsilon({\bf k}) & = & -2(t_{1,2}\cos k_a+t_{2,1}\cos k_b)-\mu
\nonumber \\
  & = & -4t\cos\frac{k_x}{\sqrt{2}}\cos\frac{k_y}{\sqrt{2}} \mp 
 4t\epsilon\sin\frac{k_x}{\sqrt{2}}\sin\frac{k_y}{\sqrt{2}}-\mu \ ,
\label{dispersion}
\end{eqnarray}
where  $t_{1,2}=t(1\pm\epsilon)$, and the lattice constant is taken to 
be unity.

Quasiparticles scatter on the twin boundary preserving their energy 
and the parallel component of their momentum. To study the scattering 
process we first find from Eq.~(\ref{dispersion}) normal components of 
the Fermi momenta for incoming and outgoing electrons with a given 
parallel component $k_y$ (see Fig.~\ref{FS}). In the right twin they 
are $k_{\pm x}^>=\pm k+q$, while in the left twin 
$k_{\pm x}^<=\pm k-q$, 
\begin{equation}
\tan\frac{q}{\sqrt{2}}=\epsilon\tan\frac{k_y}{\sqrt{2}}\ ,  \ \ \ \ 
\cos\frac{k}{\sqrt{2}}=\frac{-\mu\cos(q/\sqrt{2})}
{4t\cos(k_y/\sqrt{2})}\ .
\end{equation}
The normal components of the Fermi velocities for all these states 
have, however, the same absolute value $v_{Fx}\! =\! 
2\sqrt{2}t\sin(k/\sqrt{2})\cos(k_y/\sqrt{2})/\cos(q/\sqrt{2})$. 
(Note, that velocities are defined with dimension of energy.)

The wave function describing the scattering of an electron approaching 
the twin boundary from the right is 
\begin{equation}
\psi^>_i\! = \! e^{i{\bf k}^>_-\cdot{\bf r}_i} \! + \!  
r e^{i{\bf k}^>_+\cdot{\bf r}_i} \,(x\!\!>\!\!0), \ \ 
\psi^<_i \! = \! d e^{i{\bf k}^<_-\cdot{\bf r}_i} \,(x\!\!<\!\!0).
\end{equation}
Two equations for the unknown amplitudes $r$ and $d$ are 
(i) continuity of the wave function $\psi_0^>=\psi_0^<$ and 
(ii) Eq.~(\ref{Sch}) for $x=0$, which we write, after some algebra and 
Fourier transformation over $y$, as 
\begin{equation}
\psi^>_1e^{-iq/\sqrt{2}}-\psi^<_1e^{iq/\sqrt{2}} 
= \frac{U_0\cos(q/\sqrt{2})}{2t\cos(k_y/\sqrt{2})}\,\psi_0 \ .
\label{discrete}
\end{equation}
Remarkably, if $U_0=0$, quasiparticles move across the twinning plane 
without scattering for arbitrary degree of orthorhombicity $\epsilon$. 
The potential barrier on the boundary yields a nonzero probability of 
the reflection with amplitude $r=\alpha/(i-\alpha)$, $\alpha= 
U_0/\sqrt{2}v_{Fx}$. 
  
Below $T_c$ we again will not attempt to solve the problem 
self-consistently but rather use a ``guess'' order parameter to study 
qualitative features of the quasiparticle spectrum. The discrete BdG 
equations are 
\begin{equation}
E\psi_i=\sum_j\left(\begin{array}{cc}
t_{ij}-\mu\delta_{ij} & \Delta_{ij} \\
\Delta^*_{ij} & -t_{ij}+\mu\delta_{ij} 
\end{array}\right) \psi_j \ ,
\label{BdG}
\end{equation} 
where $\psi$ stands now for a vector with quasielectron and quasihole 
components. For orthorhombic symmetry and nearest-neighbor 
interactions we define $\Delta_{i\,i\pm a}=\Delta_1/2$, 
$\Delta_{i\,i\pm b}=-\Delta_2/2$ in one twin and 
$\Delta_{i\,i\pm a}=\Delta_2/2$, $\Delta_{i\,i\pm b}=-\Delta_1/2$ 
in the other, $\Delta_{1,2}=\Delta(1\pm\delta)$ (see Fig.~\ref{TB}).  
The superconducting gap in the two twins is given by Fourier transform 
of the off-diagonal element in (\ref{BdG}):
\begin{equation}
\Delta({\bf k}) = \Delta(\cos k_a-\cos k_b)\pm
\Delta\delta(\cos k_a+\cos k_b) \ .
\label{gap}
\end{equation}
A nonzero parameter $\delta$ describes effect of crystal 
orthorhombicity on superconducting properties and determines a 
relative weight of the $s$-wave component. 

In our model of the twin boundary the first term of Eq.~(\ref{gap}), 
i.e., the $d$-wave component, is chosen to be constant, whereas the 
second, extended $s$-wave part has a discontinues change of sign at 
the twinning plane. Although this assumption is not correct and 
$\Delta_s$ varies on a characteristic scale $\xi_s\gg a$, the 
qualitative picture remains the same since the discussed results 
depend only on the asymptotes of the order parameter. In the following 
analysis we will use the symmetry of the problem coming from relations 
$t_{\hat{\sigma}i\hat{\sigma}j}=t_{ij}$ and 
$\Delta_{\hat{\sigma}i\hat{\sigma}j}=-\Delta_{ij}$, where 
$\hat{\sigma}$ is reflection in the twinning plane. The BdG equations 
are invariant under the combined transformation 
$\hat{U}=\tau_3\hat{\sigma}$, $\tau_3$ being the Pauli matrix which 
acts in the particle-hole space. All solutions of Eqs.~(\ref{BdG}) 
are, hence, classified by their parity with respect to $\hat{U}$. 

In the bulk of each twin, Eqs.~(\ref{BdG}) are diagonalized by a 
transformation to plane waves. The wave function of a bound state is 
their linear combination
\begin{eqnarray} 
\psi^>_i\!&=&\!\!\left(\!\!\begin{array}{c}\Delta_-\\E+i\Omega_-
\end{array}\!\!\right)\! e^{i{\bf k}_-\cdot{\bf r}_i-\kappa_-x}\!+\! 
R\left(\!\!\begin{array}{c} \Delta_+\\E-i\Omega_+\end{array}\!\!\right) 
 \!e^{i{\bf k}_+\cdot{\bf r}_i-\kappa_+x}     
\nonumber \\ & & \Omega_\pm=\sqrt{\Delta_\pm^2-E^2} \ , \ \ \ 
\kappa_\pm = \Omega_\pm/v_{Fx} 
\label{bs}
\end{eqnarray}
for $x>0$, while on the left side $\psi^<_i$ for even and odd 
eigenfunctions are obtained from $\psi^<_i = 
\pm\tau_3\psi^>_{\hat{\sigma}i}$. The energy of the bound state and 
the parameter $R$ are determined from Eq.~(\ref{BdG}) for 
$i\in\text{TB}$. In the leading order in $\Delta$, that is, neglecting 
corrections $O(\Delta^2/\varepsilon_F)$ to the energy, the 
characteristic equation coincides with Eq.~(\ref{discrete}) for the 
two-component function (\ref{bs}). Solving it for $U_0=0$ we find that 
all bound states have zero energy independent of normal state 
anisotropy. They exist if the gaps of incident and outgoing 
quasiparticles satisfy 
\begin{equation}
\Delta_+\Delta_- > 0 \ ,
\label{condition}
\end{equation}
e.g., for $k_{y\text{C}}<k_y<k_{y\text{A}}$ and for $k_y > 
k_{y\text{D}}$ and $k_y<k_{y\text{B}}$ in Fig.~\ref{FS}. Taking into 
account the symmetry of the twin boundary it is easy to see that this 
condition coincides with the previously obtained 
$\Delta({\bf k}_F,\infty)\Delta({\bf k}_F,-\infty)<0$. It is, 
however, opposite to the condition on the existence of surface 
bound states in a $d$-wave superconductor \cite{Hu94}.   

The boundary potential $U_0$ splits the degeneracy between even and 
odd states. For quasiparticles moving nearly perpendicular to the 
twinning plane, the dimensionless parameter $\alpha = 
U_0/\sqrt{2}v_{Fx} \sim U_0/\varepsilon_F$ is small. Bound states 
exist in the same region as in unperturbed case (\ref{condition}) and 
their energies acquire dispersion 
\begin{equation}
E_{\text{even,odd}} = 
\pm \frac{2\alpha \Delta_+\Delta_-}{|\Delta_+ + \Delta_-|} \ ,
\label{eb}
\end{equation}
which yields a finite width of zero-energy peak in the local density 
of states (\ref{locden}). The effect of the boundary potential is 
destructive for bound states with wave vectors ${\bf k}_\pm$ nearly 
parallel to the twin boundary, i.e., for bound states with 
$k_y>k_{y\text{D}}$ and $k_y<k_{y\text{B}}$ in Fig.~\ref{FS}. A weak 
barrier $U_0/\varepsilon_F\sim\Delta_s/\Delta_0$ is sufficient to 
destroy completely these bound states. 

As a further check of the robustness of the midgap bound states, we 
have relaxed the condition of constant nearest-neighbors pairing 
amplitudes in the Hamiltonian and changed them near the twin boundary 
as shown in Fig.~\ref{TB}. This change models a space dependent gap on 
the discrete lattice and again does not affect energies of the bound 
states irrespective of the difference between $\Delta$'s and 
$\Delta'$'s. 

The transmission of excitations by the twin boundaries, which is 
important for the heat conduction process at low temperatures, also 
has unusual features. In particular, for $T<\Delta_s$ the heat is 
carried to the twin boundary by low energy excitations with ${\bf 
k}\approx{\bf k}_{\text{A}}$ in Fig.~\ref{FS}. For these 
quasiparticles $|\Delta_-|<E<|\Delta_+|\approx\Delta_s$. Let us 
consider such an electron-like state with ${\bf k}\approx{\bf k}^>_-$ 
approaching the boundary from the right. It cannot be reflected back 
into the twin as an electron with ${\bf k}\approx{\bf k}^>_+$, nor 
transmitted across the boundary as an electron with ${\bf k} \approx 
{\bf k}^<_-$, since the gap magnitude for the outgoing quasiparticles 
is $|\Delta_+|>E$. However, this electron can transform into a hole 
having ${\bf k}\approx{\bf k}^>_-$ and the same gap $\Delta_-$ but 
with the opposite group velocity (Andreev reflection \cite{Andreev}). 
In the case of zero barrier ($U_0=0$) the Andreev reflection is 
perfect (there is only an exponentially decaying wave in the left 
twin). The transmission of heat across the twin boundary must thus 
come from excitations with $E>\Delta_s$ and there will be a 
temperature dependent  contribution to the thermal conductivity from 
the twinning planes of the form $\kappa_{\text{TB}}\propto 
\exp(-\Delta_s/k_BT)$. 

An extra potential $U_0$ on the twin boundary, however, opens up an 
additional scattering channel in which an incident electron-like 
quasiparticle transforms into an outgoing hole-like excitation with 
${\bf k}\approx{\bf k}^<_+$ on the other side of the boundary 
(Fig.~\ref{FS}). It is this process of Andreev {\it transmission}, 
which will be responsible for the thermal conductivity across the twin 
boundaries in the low temperature limit. Interestingly, the boundary 
barrier $U_0$, which causes reflection of the quasiparticles from the 
twin boundary in the normal state, allows transmission of the 
low-energy excitations in the superconducting state. 

Finally, we comment on properties of bound quasiparticles for the 
time-reversal breaking superconducting state of the twin boundary 
\cite{Sigrist96}. These effects can be most easily understood on the 
basis of Eq.~(\ref{Andreev}). We assume that the gap function has, in 
addition to (\ref{ansatz}), a small imaginary component of the 
$s$-wave harmonic $i\Delta_sf(x)$, $f(x)$ being an even function. The 
right-hand side of Eq.~(\ref{Andreev}) contains in this case an extra 
term $-\Delta_s\tau_2f(x)$. First-order energy corrections found by 
averaging this perturbation with respect to the unperturbed states 
(\ref{states}) are nonzero and have opposite signs for $\psi_+$ and 
$\psi_-$. As a result, the bound state peak (\ref{locden}) in the 
local density of states shifts to finite energies, making possible to 
distinguish the two superconducting states on twinning planes in 
tunneling experiments. Condensation of bound states with a nonzero 
momentum produces also a local current flow parallel to the twin 
plane, which was derived in \cite{Sigrist96} in the framework of the 
Ginzburg-Landau theory. 

In conclusion, the formation of electron bound states on twin 
boundaries is a general property of orthorhombic predominantly 
$d$-wave superconductors. Observation of a bound state peak in the 
local density of states on twinning planes in YBCO by means of 
scanning tunneling microscopy would be a direct confirmation of the 
presence of $s$-wave component in CuO$_2$ planes. Also, a novel 
process of Andreev transmission is responsible for thermal 
conductivity across the twin boundaries in the low temperature limit. 

This work was supported by the National Science and Engineering 
Research Council of Canada.

\begin{figure}
\caption{Twin boundary and nearest-neighbor hopping and pairing
amplitudes; $1\equiv(t_1,\Delta_1/2)$, 
$\bar{1}\equiv(t_1,-\Delta_1/2)$, $1'\equiv(t_1,\Delta'_1/2)$, 
$\bar{1'}\equiv(t_1,-\Delta'_1/2)$, $2\equiv(t_2,-\Delta_2/2)$ etc.}
\label{TB}
\end{figure}

\begin{figure}
\caption{Twin boundary and orthorhombic Fermi surfaces in the two 
twins. The capital letters A, B, ... denote the positions of nodes of 
the $d\pm s$ gaps. The sign of $\Delta({\bf k})$ in the different 
regions on  the Fermi surfaces is indicated as $+$ or $-$. } 
\label{FS}
\end{figure}

\end{document}